\documentclass[letters,fleqn,usenatbib]{mnras}
\usepackage{graphicx}
\usepackage{amsmath}                
\usepackage{amssymb}                
\usepackage{color}
\usepackage{mathtools}
\usepackage{journals}
\usepackage[textwidth=1.2cm]{todonotes}
\usepackage{url}
\usepackage{times}
\usepackage[T1]{fontenc}    
\usepackage{ae,aecompl}     

\newcommand{\FigsFolder}{.}
\newcommand{\Mabs}{$\mathcal{M}_{r}^{\rm Petro}$}

\newcommand{\Mvir}{$\log(M_{\rm vir}/{\rm M_{\odot}})$}

\newcommand{\Rvir}{$r_{\rm vir}$}
\newcommand{\Ma}{$\log (M_{\rm a} / {\rm M}_{\odot})$}
\newcommand{\Ms}{$\log (M_{\rm s} / {\rm M}_{\odot})$}

\title[Group density profiles far beyond $r_{\rm vir}$]
{Group galaxy number density profiles far out: is the `one-halo' term NFW out to $>10$ virial radii?}
\author[Trevisan, Mamon \& Stalder]{M. Trevisan$^{1}$\thanks{E-mail:
trevisan@iap.fr}, G. A. Mamon$^{1}$, D. H. Stalder$^{1,2}$ \\
$^{1}$Institut d'Astrophysique de Paris (UMR 7095: CNRS \& UPMC, Sorbonne-Universit\'es), 98 bis Bd Arago, 75014 Paris, France \\
$^{2}$Instituto Nacional de Pesquisas Espaciais/MCT, Av. dos Astronautas 1758, 12227-010
S\~ao Jos\'e dos Campos, Brazil}

\begin{document}

\date{Accepted ---. Received ---; in original form ---}

\pagerange{\pageref{firstpage}--\pageref{lastpage}} \pubyear{2017}

\maketitle

\label{firstpage}

\begin{abstract}
While the density profiles (DPs) of $\Lambda$CDM haloes obey the NFW law out 
to roughly one virial radius, \Rvir, the structure of their outer parts is
still poorly understood, because the 1-halo term describing the halo itself
is dominated by the 2-halo term representing the other halos picked up. 
Using a semi-analytical model,
we measure the real-space 1-halo number DP of groups out to 20 \Rvir\  
by assigning each galaxy to its nearest group above mass $M_{\rm a}$, 
in units of the group \Rvir.
If $M_{\rm a}$ is small (large), the
outer DP of groups falls rapidly (slowly). 
We find that  there is an optimal $M_{\rm a}$ for which the stacked
DP resembles the NFW model to 0.1 dex accuracy out to
13 virial radii. 
We find similar long-range NFW surface DPs (out to 10 \Rvir) 
in the SDSS observations using a galaxy assignment
scheme that combines the non-linear virialized regions of groups with their linear
outer parts. 
The optimal $M_{\rm a}$ scales as the minimum mass of the groups
that are stacked to a power 0.25--0.3. 
The NFW model thus does not solely originate from violent
relaxation.
Moreover,  populating haloes with galaxies using HOD
models must proceed out to much larger radii than usually done. 
\end{abstract}


\section{Introduction}

Cosmological dissipationless N-body simulations have taught us that,
regardless of their mass,
the radial density profiles (DPs) of haloes in the range of
$\approx$ 0.01 to 1.5 virial radii ($r_{\rm vir}$)
are well described by the \citet*[][NFW]{Navarro.etal:1996} model whose inner and outer slopes are respectively $-1$ and $-3$
(\citealt{Navarro.etal:2004} find that 
the \citealt{Einasto:1969} model provides an even better representation of
the DP, with a more progressive change of slopes).

The origin of the NFW profile could be a combination of fast and slow accretion in the
inner and outer region, respectively (e.g., \citealt{Lu.etal:2006}), where the fast
accretion is generally related to violent relaxation.
The total DP is understood to be the sum of two terms (as first introduced by \citealt{Cooray.Sheth:2002} in the
context of galaxy clustering):
the $\hbox{1-halo}$ term describing the halo itself, and the $\hbox{2-halo}$ term describing
the other haloes around the first one, following the $\hbox{2-point}$ correlation
function of haloes. 
Beyond a few $r_{\rm vir}$, the 1- and 2-halo
terms respectively correspond to an extension of the halo and the other haloes
outside. While several authors studied the sum of both terms 
beyond $r_{\rm vir}$ (e.g., \citeauthor{Prada.etal:2006}; \citealt{Hayashi.White:2008,Diemer.Kravtsov:2014}), 
they all assumed possibly truncated NFW or Einasto profiles 
for the 1-halo term (and their stacked DPs
involved multiple counting).
Hence, the 1-halo term is poorly known beyond $\approx\,2\,r_{\rm vir}$.

In this \emph{Letter}, we assign galaxies in a  semi-analytical model (SAM)
of galaxy formation to their nearest group in units of the group's $r_{\rm
  vir}$. This allows us 
to explore the $\hbox{1-halo}$ term by measuring the DPs
of groups traced by their galaxies out to $20\,r_{\rm vir}$.
We then compare these DPs to the \emph{galaxy surface number
density profiles} (SDPs) of groups in the Sloan Digital Sky Survey (SDSS), also out to $20\,r_{\rm vir}$,
using a novel scheme to assign each galaxy to its closest group in redshift space.  
In Sect.~\ref{Sec_data}, we describe the simulation and data used.  
Our assignment scheme is explained in Sect.~\ref{Sec_assign} and in Sect.~\ref{Sec_results} we present the results of our study. 
Finally, we summarize and discuss our results in Sect.~\ref{Sec_summary}.
Masses and distances are given in physical units, and
 we adopt the $\Lambda$CDM cosmology with $\Omega_{\rm m} = 0.275$, $\Omega_{\Lambda} = 0.725$, and
${H_0 = 70.2~\hbox{km s}^{-1} \hbox{Mpc}^{-1}}$ \citep[WMAP7,][]{Komatsu.etal:2011}.

\section{Observations and simulation}
\label{Sec_data}

\subsection{SDSS galaxies and groups}
\label{Sec_SDSS}

The observational sample of galaxies was retrieved from the SDSS-DR12
\citep{Alam.etal:2015} database. We selected all galaxies from the Main
Galaxy Sample that
are in the redshift range $0.01 \leq z \leq 0.05$ and are more luminous
than $\mathcal{M}^{\rm Petro}_r \leq -18.78$, where \Mabs\ corresponds
to the k-corrected absolute Petrosian magnitude in the $r$-band. These
criteria lead to a doubly-complete subsample in distance and luminosity containing $63,642$ galaxies.
The k-corrections were computed with the {\sc kcorrect} code (version 4\_2)
of \citet{Blanton.Roweis:2007}, and we obtained the magnitude limit of the
sample using a geometric approach 
similar to that of  \citet*{Garilli.etal:1999}.

The galaxy groups were selected from the updated version of the catalogue compiled by \citet{Yang.etal:2007}\footnote{We used 
the catalogue {\tt petroB}, which is available at \url{http://gax.shao.ac.cn/data/Group.html}.}. 
The new catalogue contains 473,482 groups drawn from a sample of 601,751 galaxies mostly from the 
Sloan Digital Sky Survey's Data Release~7 \citep[SDSS-DR7,][]{Abazajian.etal:2009}.

The radii $r_{200,{\rm m}}$ (of spheres that are 200 times denser than the 
\emph{mean} density of the Universe) are derived from the $M_{200,{\rm m}}$ masses given in the \citeauthor{Yang.etal:2007}
catalogue, which are based on abundance matching with the group luminosities. 
We then determined the virial radii, $r_{\rm vir}$, 
the corresponding virial masses, $M_{\rm vir} = (\Delta_{\rm v}/2)\, H^2(z)\,r_{\rm
  vir}^3/G$, and virial velocities $v_{\rm vir} = \sqrt{\Delta_{\rm v}/2}\,H(z)\,r_{\rm
  vir}$,
defined such that the mean densities within the virial sphere are $\Delta_{\rm v}$=100 times the \emph{critical} density of the
Universe,\footnote{See appendix A in \citet*{Trevisan.etal:2017a} for the conversion from quantities relative to the 
mean density to those relative to the critical density.} by assuming the 
NFW DP and the concentration-mass relation of 
\cite{Dutton.Maccio:2014}. 

To avoid incomplete profiles of SDSS groups, we first assure that at least $95\%$ of the region within ${20\, r_{\rm vir}}$ from the group centres lies
within the SDSS coverage area. 
For this purpose, we adopted the SDSS-DR7 spectroscopic angular selection function mask\footnote{We used the file {\tt sdss\_dr72safe0\_res6d.pol}, 
which can be downloaded from \url{http://space.mit.edu/~molly/mangle/download/data.html}} provided by the NYU Value-Added
Galaxy Catalog team \citep{Blanton.etal:2005} and assembled with the package {\sc Mangle 2.1} \citep{Hamilton.Tegmark:2004, Swanson.etal:2008}.
We also require that the groups lie far enough from the redshift limits of
the galaxy sample ($z_{\rm min} = 0.01$ and $z_{\rm max} = 0.05$), by
only selecting groups within the redshift range 
$[{z_{\rm min} + 20\, \Delta z,
z_{\rm max} - 20\, \Delta z}]$,
where ${\Delta z = {\sqrt{{2}/{\Delta_{\rm v}}}\,(1 + z_{\rm group})}\, v_{\rm vir}/c}$, where $c$ is the speed of light (see Sect.~\ref{Sec_assign}).
These criteria lead to a sample of $1961$ groups with \Mvir~$\ge 12.5$.

\subsection{Simulations}
\label{Sec_sims}

We used the SAM by \citet{Henriques.etal:2015}, 
which was run on the Millennium-II simulations \citep{BoylanKolchin.etal:2009}.
We extracted the snapshot corresponding to $z = 0$ from the {\tt Henriques2015a..MRIIscPlanck1} table in the Virgo--Millennium database of the 
German Astrophysical Virtual Observatory (GAVO\footnote{\url{http://gavo.mpa-garching.mpg.de/portal/}}).

From the simulation box extracted from GAVO, we built a mock flux-limited,
SDSS-like sample of groups and galaxies. Since 
the simulation box is not large enough to produce the SDSS-like group catalogue, 
we replicated the simulation box along the three Cartesian coordinates, then
placed an observer at some position and mapped the galaxies on the sky.
The absolute magnitudes in the 
$r$-band (including internal dust extinction) were converted to apparent magnitudes, and the flux limit of the
Main Galaxy Sample of the SDSS, ${m_r < 17.77}$, was applied. 

We then select the galaxies and groups from the mock catalogue
following the same selection criteria that is applied to the observations 
and presented in Sect.~\ref{Sec_SDSS}.
In particular, the doubly complete mock subsample, again limited to luminosities
$\mathcal{M}^{\rm Petro}_r \leq -18.78$, contains 61,915 galaxies.
We apply the SDSS spectroscopic mask to the mock data.

\section{Membership assignment scheme}
\label{Sec_assign}

To assign galaxies to the group that attracts them the most, 
one requires selecting the group with the lowest
distance to the group in units of virial radius, $d$  (since acceleration
decreases with distance in all  models with density slopes steeper than --1
everywhere).
This is straightforward in our real-space (3D) sample.
In our redshift-space (2+1D) sample, for galaxies far away from the group, we estimate $d$
using the standard redshift-space distance
\begin{equation}
 d_{\rm outer}(R, \Delta z) = \sqrt{\frac{\Delta_{\rm v}}{2} \left[ \frac{c \,\Delta z}{v_{\rm vir} (1 + z_{\rm group})} \right]^2 +
 \left( \frac{R}{r_{\rm vir}} \right)^2}\ .
\label{Eq_dOuter}
\end{equation}
For a galaxy lying close to a group, we take into account the strong redshift distortions by applying the
overdensity in projected phase space (PPS), $P_M(R,\Delta z)$ introduced
by \cite{Yang.etal:2005,Yang.etal:2007}, which is the suitably scaled product
of the NFW  SDP times a Gaussian distribution of
galaxy-group redshift differences.  
We convert this overdensity to an equivalent redshift-space distance by
joining the two estimators at a fixed number of virial radii, $\widetilde R_{\rm n}$,  
marking the transition from the non-linear group to the linear outer regions. This amounts to 
\begin{equation}
 d_{\rm inner}(R,\Delta z) = \left( \frac{a - \ln P_M(R,\Delta z)}{b}
 \right)^{1/2}, 
\label{Eq_dInner}
\end{equation}
\noindent where $b=1/(\eta^2\,\Delta_{\rm v})$ while $a$ is given by
\begin{equation} 
a = 
\frac{\widetilde{R}_n^2}{\eta^2\,\Delta_{\rm v}} 
+ \ln \left(
 \frac{2}{3}\,    \sqrt{\frac{\Delta_{\rm v}}{\pi}}\, \frac{H(z)}{H_0} 
    \frac{c_{\rm v}^2\, g(c_{\rm v})}{\Omega_{\rm M}\, \eta\, (1 + z)}
 f(\widetilde{R}_n) \right) \ .
\label{Eq_ab_PM}
\end{equation}
In eq.~(\ref{Eq_ab_PM}), 
$c_{\rm v}$ is the concentration parameter,
${1/g(c_{\rm v}) = \ln(1+c_{\rm v})-c_{\rm v}/(1+c_{\rm v})}$, and 
\[
f(\widetilde{R}) = {2\pi\,\Sigma(R)\,r_{\rm vir}^2\over c_{\rm v}^2\,g\!\left(c_{\rm
    v}\right)\,N_{\rm vir}} =
\frac{1 - |c_{\rm v}^2 \tilde{R}^2 - 1|^{-1/2} \ C^{-1}[1/(c_{\rm v} \tilde{R})]}{c_{\rm v}^2 \tilde{R}^2 - 1}\ ,
\]
where $C^{-1}(x) = \cos^{-1} x$ or $\cosh^{-1} x$, depending on whether $x<1$
or $x>1$. Analyzing the galaxy assignments from the 3D SAM projected into the PPS, we deduce that
$\widetilde R_{\rm n} = 2.5$.
A more detailed description of our approach, as well as the full derivation of eqs.~(\ref{Eq_dInner}) and (\ref{Eq_ab_PM}) 
are given in a forthcoming paper (Trevisan et al.~2017, in prep.)

\subsection{Group mass thresholds for the assignment}
\label{Sec_masses}

We consider two group mass thresholds. The first one, $M_{\rm s}$,
corresponds to the minimum virial mass of the groups in our \emph{sample} around which we are
measuring the number DPs.
The second, $M_{\rm a}$, is the lowest group mass to which we can \emph{assign}
galaxies.. 
When $M_{\rm a}$ is extremely low,
most galaxies outside the virial radius of a group are assigned to their
one-galaxy groups, leaving few galaxies beyond that radius. On the
other hand, if $M_{\rm a}$ is large, we partially pick up the 2-halo term
in our group DP.

\section{Results}
\label{Sec_results}

\subsection{Three-dimensional number density profiles}
\label{Sec_3Dprof}

Fig.~\ref{Fig_density_3d} shows the galaxy number DPs obtained in the
simulations for stacked groups from the SAM with \Mvir~$> 13.0$, using 3 
different values of $M_{\rm a}$.
We fit the 
parameters of the NFW and Einasto models using maximum likelihood estimation (MLE); therefore, no binning of the data is required.
The MLE was performed considering only the galaxies within the region $0.1 \leq r/r_{\rm vir} \leq 2$. 

The middle and bottom panels in Fig.~\ref{Fig_density_3d} shows the residuals of the best-fit profiles. 
For $\log (M_{\rm a}/{\rm M}_{\odot}) = 12.3$, the NFW describes the density
profile very well to 0.1 dex accuracy out to $r \sim 13~r_{\rm vir}$. 
On the other hand, the Einasto form fails to describe the DP in the outer regions, as shown in the bottom panel in Fig.~\ref{Fig_density_3d}. 
A reasonable fit requires  including the outer regions in the MLE procedure. 
For that model,  fitting the profile between $0.1 \leq r/r_{\rm vir} \leq 8$
leads to 0.1 dex residuals from
$r \sim 0.2$~\Rvir\ to $\sim 13\,r_{\rm vir}$, with best-fit parameters $c_{\rm v} = 5.5 \pm 0.8$ and $n = 8.9 \pm 0.6$.

\subsection{Surface density profiles and comparison with observations}
\label{Sec_prof}

Applying the method described in Sect.~\ref{Sec_assign} to the SDSS data, we obtain the SDP shown in Fig.~\ref{Fig_density}. 
Since our scheme is designed to assign galaxies within a sphere of radius 
20~\Rvir, 
we compare the SDP with that of the NFW model computed by
integrating the 3D DP along the line-of-sight within that sphere
(its analytical form is provided in appendix B.1 of \citealt*{Mamon.etal:2010a}).

\begin{figure}
\includegraphics[width=\hsize]{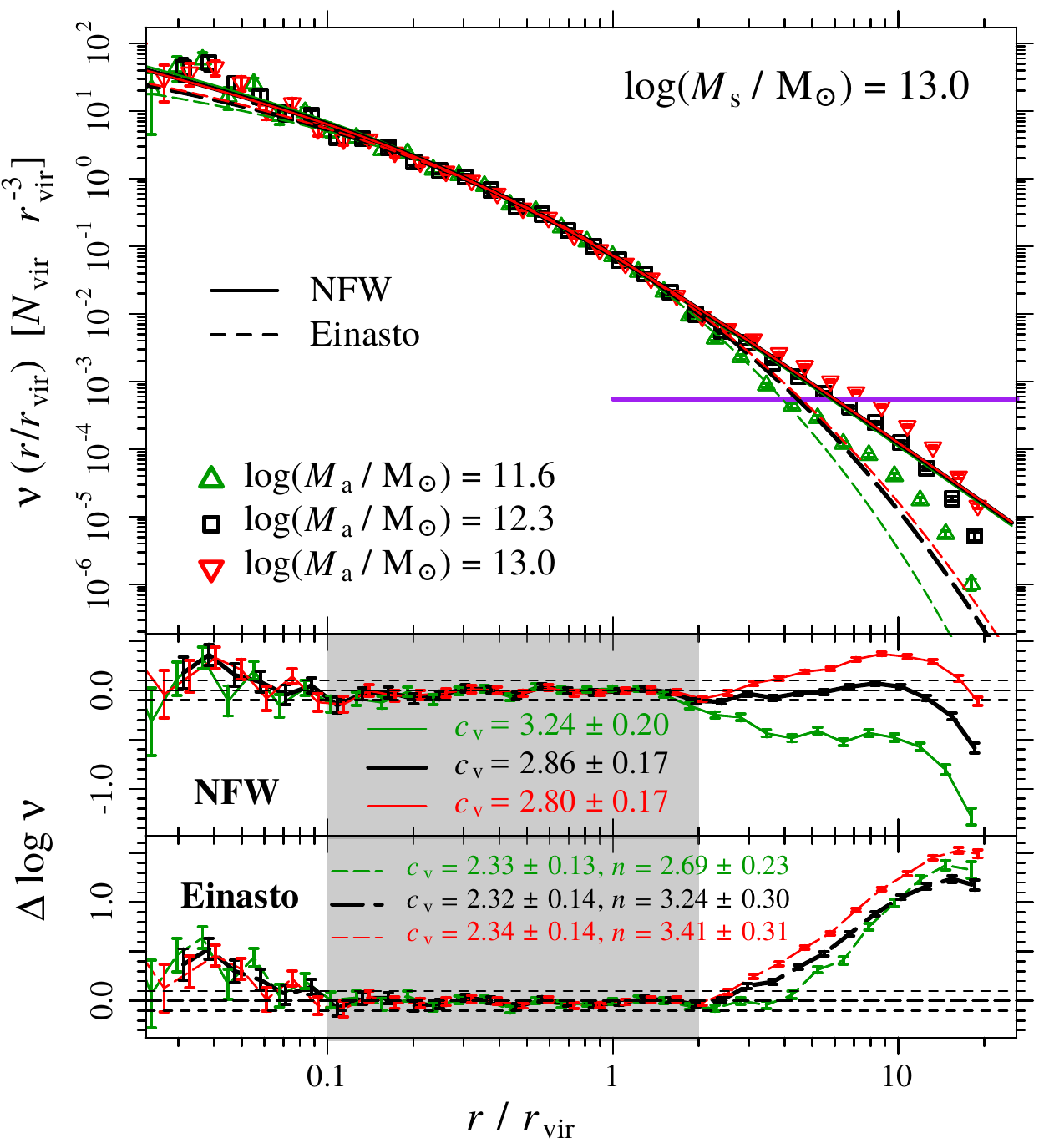}
 \vspace{-0.3cm}
\caption {Number density profile of 525 stacked groups with $\log(M_{\rm s}/\rm M_\odot) = 13.0$ from the simulation. 
Galaxies were assigned to groups of minimum mass as indicated.
The mass threshold
${\log (M_{\rm a} / {\rm M}_{\odot}) = 12.3}$ (black) 
leads to the density profile that 
is best described by the NFW model out to ${r \sim 13~r_{\rm vir}}$ (see Sect.~\ref{Sec_Ma}), with 
residuals of $\pm 0.1$~dex (\emph{middle panel}).
The \emph{purple horizontal line} represent the mean density of the Universe.
The \emph{middle} and \emph{bottom panels} show the residuals of the best-fit NFW and Einasto profiles. 
The \emph{shaded areas} indicate the region considered in the 
fitting procedure ($0.1 < r/r_{\rm vir} < 2$), 
and the \emph{long} and \emph{short-dashed horizontal lines} respectively indicate
$\Delta \log \nu = 0$ and $\pm 0.1$~dex.
The colours are the same as in the upper panel, and the 
best-fit parameters are indicated in each panel. 
The errors in the data points are from 1000 bootstraps on the groups combined with
Poisson, while those on parameters $c_{\rm v}$ and $n$ are from those bootstraps.}
\label{Fig_density_3d}
\end{figure}

\begin{figure}
\centering
\includegraphics[width=\hsize]{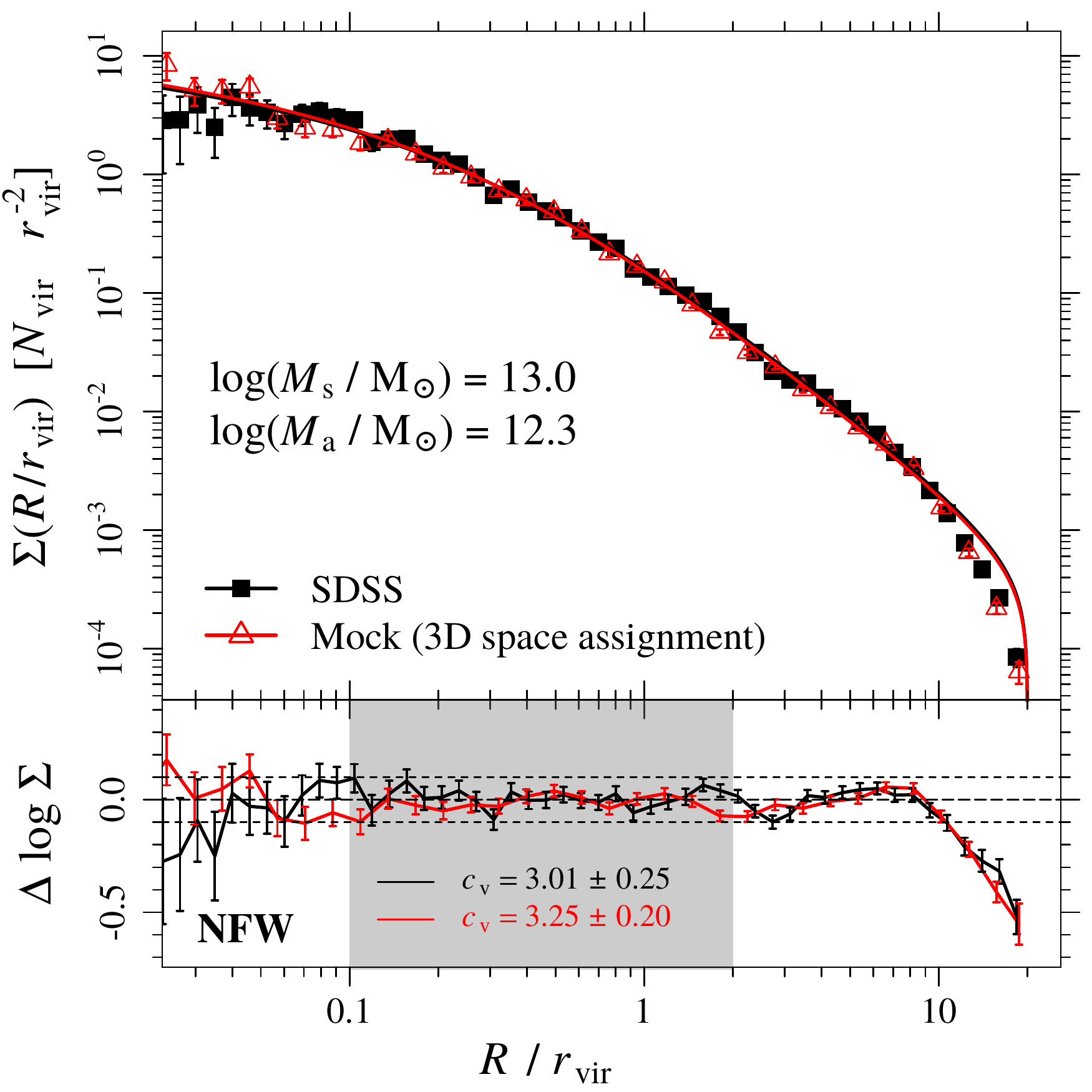} 
 \vspace{-0.3cm}
\caption {Surface number density profiles of stacked groups from the simulations (\emph{red}, 525 groups) and SDSS data (\emph{black}, 534 groups). 
The \emph{upper panel} shows the stacked profile of groups with \Mvir~$\ge 13.0$ obtained for ${\log (M_{\rm a}/{\rm M}_{\odot}) = 12.3}$.
The profile from the simulations is the projection of the profile shown in Fig.~\ref{Fig_density_3d}.
The residuals of the best-fit NFW profiles limited to the $20$~\Rvir\ sphere are shown in the \emph{bottom panel}, where the  
\emph{shaded area} indicates the region
where the fit is performed (${0.1 < R/r_{\rm vir} < 2}$). 
The \emph{long} and \emph{short-dashed horizontal lines} indicate $\Delta \log \Sigma = 0$ and $\pm 0.1$~dex, respectively.
The colours are the same as in the upper panel, and the concentration
parameters $c_{\rm v}$ of the best fits are presented. 
The error bars are as in Fig.~\ref{Fig_density_3d}.}
\label{Fig_density}
\end{figure}

In Fig.~\ref{Fig_density}, 
the observed profile is also compared to the projection of the 3D profile shown in Fig.~\ref{Fig_density_3d}.
The excellent agreement between the profile of SDSS groups and the simulation can be clearly seen, and 
the difference between the the best-fit $c_{\rm v}$ values are within the errors. 
This indicates that our scheme for distances in redshift-space for the SDSS sample is a good
approximation to the 3D space assignment.

\subsection{Group mass thresholds}
\label{Sec_Ma}

In Sects.~\ref{Sec_3Dprof} and \ref{Sec_prof}, we showed that the NFW profile is a good description of both the simulation and the SDSS data for groups 
more massive than \Ms~$=13$ when \Ma~$= 12.3$. However, does this result still hold for different values of $M_{\rm s}$ and $M_{\rm a}$? 

To tackle this question, we considered different group samples with $M_{\rm
  vir} \ge M_{\rm s}$ with
\Ms\ ranging from $12.5$ to $14$, in steps of $0.1$~dex. 
For each of these samples, we fit the NFW profile in the region from $0.1$ to $2$~\Rvir, and,
from the extrapolation of the best fit, we compute the predicted number of galaxies within the region from $2$ to $10$~\Rvir, $N^{\rm NFW}_{\rm outer}$.
We then determine the the best value of $M_{\rm a}$ for each $M_{\rm s}$, which corresponds to the one leading to $\Delta N = N^{\rm NFW}_{\rm outer} - N^{\rm obs}_{\rm outer} = 0$,
where $N^{\rm obs}_{\rm outer}$ is the observed number of galaxies within the region $2$ to $10$~\Rvir. 

\begin{figure}
\centering
\includegraphics[width=\hsize]{\FigsFolder/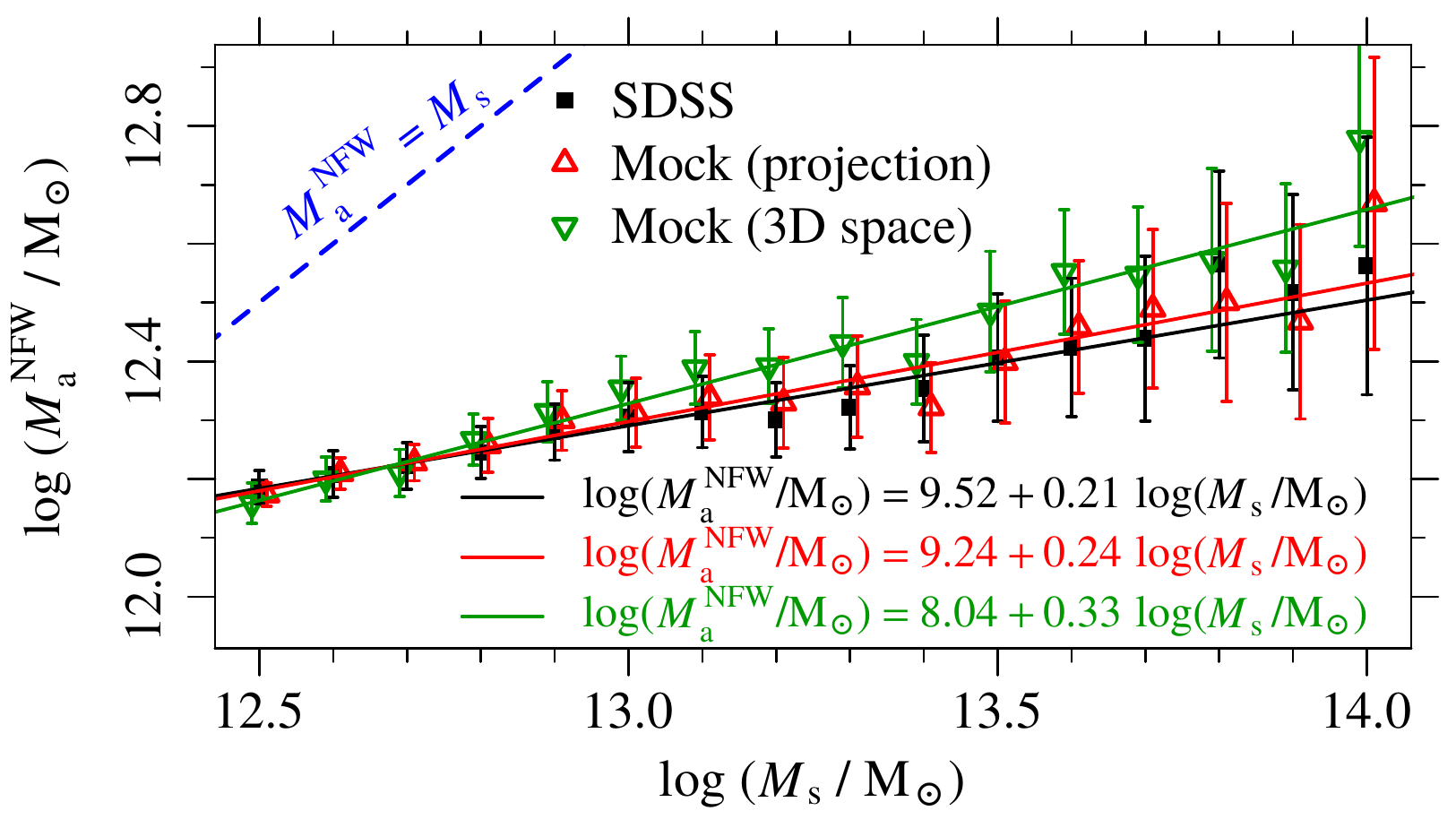}
 \vspace{-0.4cm}
\caption {Best-fit group mass threshold $M_{\rm a}^{\rm NFW}$ to obtain NFW density
  profiles out to $10\,r_{\rm vir}$
versus the minimum sample group mass $M_{\rm s}$. 
The plot shows the results from the SDSS data (\emph{black symbols}),
 the simulations in 3D (\emph{green}), and in projection (\emph{red}). 
 The errors in $M_{\rm a}^{\rm NFW}$ were estimated by bootstrapping 
the groups of each sample $200$ times, and the best linear fits to the 
 $\log M_{\rm a}^{\rm NFW}$ versus $\log M_{\rm s}$ relation are shown as
 \emph{solid lines}, while
$M_{\rm a}^{\rm NFW} = M_{\rm s}$ is shown as a \emph{dashed blue line}. 
}
\label{Fig_logMa}
\end{figure}

The resulting values of $M_{\rm a}$ that provide the closest match to the NFW
model up to very large radii (hereafter, $M_{\rm a}^{\rm NFW}$)
are shown as a function of $M_{\rm s}$ in Fig.~\ref{Fig_logMa}. For
\Ms\ ranging from $12.5$ to $14$, the values of $\log(M_{\rm a}^{\rm NFW}/\rm M_\odot)$ 
vary from $\sim$12.2 to $\sim$12.8. 
We fitted  $M_{\rm a}^{\rm NFW}$ as a function of $M_{\rm s}$ with a linear relation
${\log M_{\rm a}^{\rm NFW} = \alpha + \beta \log M_{\rm s}}$, finding
\[
\!\!\!\!
\begin{array}{l}
\alpha  = 9.52 \pm 0.16 \,,\  \beta = 0.21 \pm 0.01\ ({\rm SDSS}), \\
\alpha  = 9.24 \pm 0.18 \,,\  \beta = 0.24 \pm 0.01\ ({\rm mock\ in\ projection}), \\
\alpha  = 8.04 \pm 0.22 \,,\  \beta = 0.33 \pm 0.02\ ({\rm mock\ in\ 3D\ space}). \\
\end{array}
\!\!
\label{Eq_ab_Ma}
\]

For \Ms~$\gtrsim 13.0$, the  values of $M_{\rm a}^{\rm NFW}$ for the 3D
profiles are slightly higher than those for the projected profiles, because  
we minimise $\Delta N$ within $r = 2$ to $10$~\Rvir, which does not exactly correspond to the same range in projection.

\begin{figure*}
\centering
\includegraphics[width=\hsize]{\FigsFolder/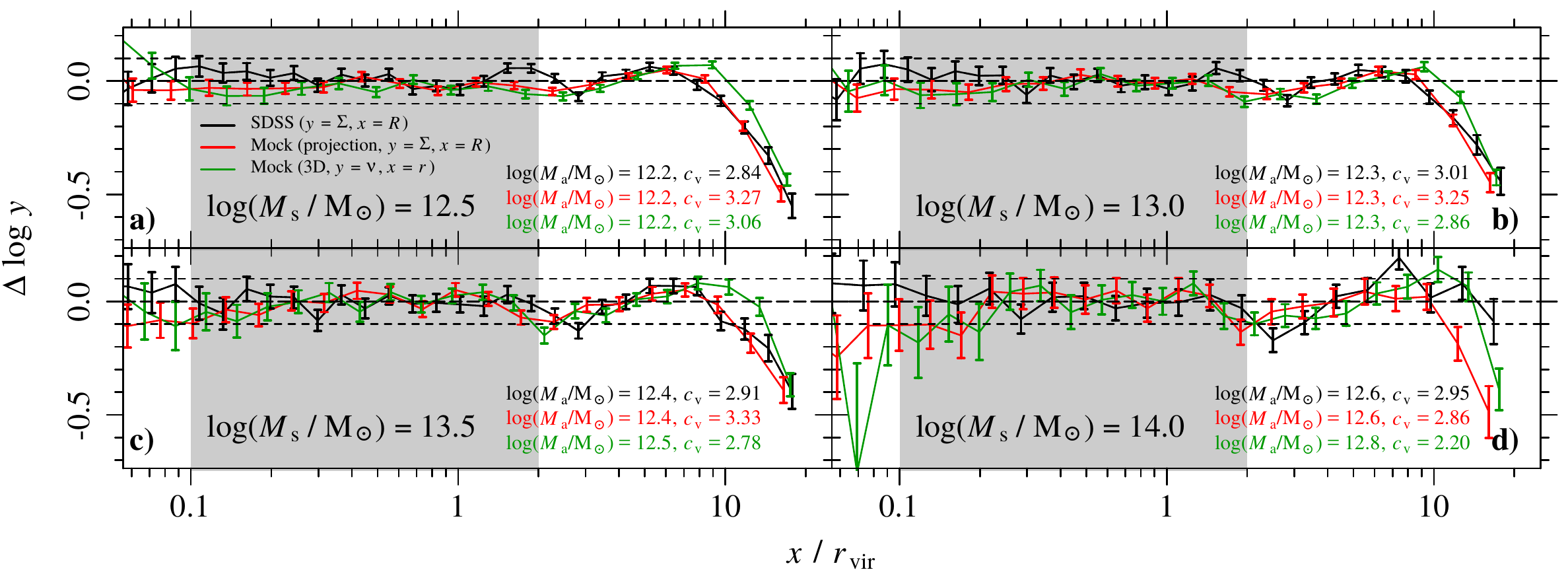} \\
 \vspace{-0.2cm}
\caption {Residuals of the density profiles relative to the best NFW fit, for four values of $M_{\rm s}$. 
The curves correspond to the results for the SDSS data (\emph{black lines}), 
the simulations in projection (\emph{red}) and in 3D (\emph{green}). 
For the 3D number density profile, 
$\Delta \equiv \Delta \nu$ and $x \equiv r$, the 3D radial distance to
the group centre. For the projected profiles (black and red lines), $\Delta
\equiv \Delta \Sigma$ 
and $x\equiv R$, the projected
distance to the group centre. The \emph{shaded areas} indicate the region of the 
fitting procedure ($0.1 < x/r_{\rm vir} < 2$).
The values of $M_{\rm a}$ correspond 
to the optimal values shown in Fig.~\ref{Fig_logMa}. 
The error bars are as in Fig.~\ref{Fig_density_3d}.}
\label{Fig_res}
\end{figure*}

The residuals of the best fits of \Ma\ for the cases when \Ms~$=12.5$, $13.0$, $13.5$, and $14.0$ are presented in Fig.~\ref{Fig_res}. 
In all cases, the DP is matched to within 0.1 dex by the NFW model out to $\sim 12$ to $14$~$r_{\rm vir}$ in real space and $10\,r_{\rm
  vir}$ in projection.

\section{Conclusions and Discussion}
\label{Sec_summary}

In this \emph{Letter}, we investigated the 1-halo term of the galaxy number DPs of groups and clusters
out to $20$ virial radii, analyzing both a recent state-of-the-art semi-analytical model of galaxy formation 
based on the Millennium-II simulations, as well as  a complete sample of
galaxies in and around groups and clusters from SDSS-DR7.  
We assigned galaxies to the nearest group in units of that group's virial
radius, which is straightforward in 3D.
In 2+1D, we use a scheme to estimate 3D distances by combining 
the non-linear behaviour within $2.5\, r_{\rm vir}$ and redshift space
distances beyond.
Our assignment method involves two group mass thresholds: 
the minimum group mass in our sample, $M_{\rm s}$, and the minimum group mass,
$M_{\rm a}$, to which we assign galaxies. 

Our main findings can be summarized as follows:
\vspace{-2mm}
\begin{itemize}
\itemsep 0pt
 \item [--] For \Ma~$=12.3$, the NFW formula describes very well
the density profile of \Mvir~$\ge 13.0$ groups  out to far beyond the virial radius,
 for both the simulations (Fig.~\ref{Fig_density_3d}) and observations (Fig.~\ref{Fig_density}). 
 Our best NFW fit, performed in the range 0.1 to $2\,r_{\rm
   vir}$ (where the NFW model is known to fit the galaxy distribution,
 \citealt{Carlberg.etal:1997}), has residuals of 0.1 dex out to distances as
 large as ${r \sim 13~r_{\rm vir}}$ (where the density is one-tenth of the
 mean density of the Universe) and projected distances $R \sim 10\, r_{\rm vir}$. 
Our best-fit concentrations for SDSS ($c_{\rm v}=3.0$) are close to the mean of the values (corrected to
our definition of $r_{\rm vir}$) of 
$c_{\rm v}=5.1$ \citep{Carlberg.etal:1997},
$c_{\rm v}=4.0$ \citep{Lin.etal:2004},
and
$c_{\rm v}=2.6$ \citep{Collister.Lahav:2005}.
 
 \item [--] On the other hand, the Einasto formula fails to describe the density profile if the best-fit parameters are estimated using only galaxies in the inner regions
 ($0.1 < R/r_{\rm vir} < 2$, Fig.~\ref{Fig_density_3d}). A good fit is obtained only when the outer regions are included in the fitting procedure.  

\item [--] For all values of \Ms\ between 12.5 and 14, i.e., from small groups to clusters of galaxies, 
we are always able to find a value of $M_{\rm a} = M_{\rm a}^{\rm NFW}$ (Fig.~\ref{Fig_res}),
ranging from $\sim 10^{12.2}$ to $10^{12.8}~{\rm M}_{\odot}$ (with $\log
M_{\rm a}$ varying linearly with $\log M_{\rm s}$, Fig.~\ref{Fig_logMa}), that leads to profiles that are
very well described by the NFW law out to $\ga 10\,r_{\rm vir}$ (even if the Einasto law can also lead to
good fits).
\end{itemize}
\vspace{-2mm}
When $M_{\rm a}$ is large, our measurement of the 
density profile  is increasingly contaminated by the ${\hbox{2-halo}}$ term
at increasingly large distances (it appears concave in
log-log).
When $M_{\rm a}$ is very low, most of the galaxies  beyond a
few virial radii are assigned to very low mass, often single-galaxy, haloes,
leaving a truncated density profile (convex in log-log).
There is, therefore, an intermediate value of $M_{\rm a}$ that represents the
transition between these two regimes. 

However, it was not obvious that intermediate 
values of $M_{\rm a}$ would lead to density profiles (both in 3D and in projection) 
that 1) are consistent with the $-3$ outer slope of the NFW model as far out as
10 virial radii and 2) are the extrapolation of  the NFW model fit only up to
$2\,r_{\rm vir}$. This suggests that 
 one could re-define the 1-halo
term as that for which the outer density profile of singly-assigned objects
(galaxies in groups) follows an NFW model.

It is intriguing  that the DPs of groups
  appear NFW-like out to $10\,r_{\rm vir}$ for the appropriate
choice of $M_{\rm a}$.
Admittedly, the origin of the power-law relation between the optimal $M_{\rm
  a}$ and $M_{\rm s}$ remains to be clarified.
Nevertheless, if this NFW behaviour at large distances for optimal values of $M_{\rm a}$ is not fortuitous,
 the origin of the outer part of the NFW model would be more complex than
previously thought. Beyond $4\,r_{\rm vir}$, a galaxy is expanding away from
its nearest group, 
but is decelerated by this group.
So, the outer $-3$ slope of the NFW model may have to do with the
combination of the primordial density field with the spherical collapse model
instead of halo mergers or slow accretion.

We observe `V'-shape kinks in the mock 1-halo DPs and SDPs at $\simeq 2.2 \,r_{\rm
  vir}$ and at $\simeq 2.8\,r_{\rm vir}$ in the SDSS 1-halo SDPs
(Fig.~\ref{Fig_res}), similar to those discovered in the total DPs \citep{Diemer.Kravtsov:2014}
and total SDSS SDP \citep{More.etal:2016}.
The presence of these kinks in our 1-halo DPs and SDPs
indicates that these are natural features of the 1-halo term related to the 2nd apocentre of orbits
(backsplash radius).

Many teams have been creating virtual galaxy catalogues by populating
galaxies in haloes using Halo Occupation Distribution models, and
nearly all truncate their galaxy distributions at
 $r \simeq r_{\rm vir}$. Our results indicate that one should instead populate
haloes with galaxies out to $\simeq 13 \,r_{\rm vir}$ of until one reaches
the next nearest group.

If the primordial density field drives the  density profiles of groups at
such large distances, one may wonder whether it also is responsible for the
variation of galaxy properties such as the increasing fraction of star forming galaxies
up to $\approx 8\,r_{\rm vir}$  observed by  \cite{vonderLinden.etal:2010}.
We investigate this question in Trevisan et al. (2017, in prep.).

\section*{Acknowledgments}

We thank the referee for comments that led to a clearer manuscript and Cristiano De
Boni for a useful comment.
MT acknowledges financial support from CNPq (process $\#204870/2014-3$).
This research has been supported in part by the Balzan foundation via the
Institut d'Astrophysique de Paris.
DHS acknowledges financial support from CNPq scholarship $\#140913/2013-0$.
We acknowledge the use of SDSS data
(\url{http://www.sdss.org/collaboration/credits.html}) and the Virgo--Millennium database 
(\url{http://gavo.mpa-garching.mpg.de/portal/}).


\label{lastpage}

\end{document}